\documentclass[runningheads]{llncs}
\usepackage{comment}
\usepackage{graphicx}
\usepackage{epsfig}
\usepackage{amsmath}
\usepackage{amssymb}
\usepackage{multirow}
\usepackage{cite}
\usepackage{xcolor}
\usepackage{url}
\usepackage{array}

\newcolumntype{P}[1]{>{\centering\arraybackslash}p{#1}}

\begin{document}
\title{Multimodal Representation Learning via Maximization of Local Mutual Information}
\titlerunning{Representation Learning via Local Mutual Information}
\author{Ruizhi Liao$^{1}$, Daniel Moyer$^{1}$, Miriam Cha$^{2}$, Keegan Quigley$^{2}$, \\
Seth Berkowitz$^{3}$, Steven Horng$^{3}$, Polina Golland$^{1}$, and William M. Wells$^{1,4}$}
\authorrunning{R. Liao et al.}
\institute{$^{1}$ CSAIL, Massachusetts Institute of Technology, Cambridge, MA, USA\\
$^{2}$ MIT Lincoln Laboratory, Lexington, MA, USA\\
$^{3}$ Beth Israel Deaconess Medical Center, Harvard Medical School, Boston, MA, USA\\
$^{4}$ Brigham and Women's Hospital, Harvard Medical School, Boston, MA, USA}
\maketitle              % typeset the header of the contribution
\begin{abstract}
We propose and demonstrate a representation learning approach by maximizing the mutual information between local features of images and text. The goal of this approach is to learn \textit{useful} image representations by taking advantage of the rich information contained in the free text that describes the findings in the image. Our method trains image and text encoders by encouraging the resulting representations to exhibit high local mutual information. We make use of recent advances in mutual information estimation with neural network discriminators. We argue that the sum of local mutual information is typically a lower bound on the global mutual information. Our experimental results in the downstream image classification tasks demonstrate the advantages of using local features for image-text representation learning. Our code is available at: \url{https://github.com/RayRuizhiLiao/mutual_info_img_txt}.

\end{abstract}
\section{Introduction}
\label{sec:intro}
We present a novel approach for image-text representation learning by maximizing the mutual information between local features of the images and the text. In the context of medical imaging, the images could be, for example, radiographs and the text could be radiology reports that capture radiologists' impressions of the images. A large number of such image-text pairs are generated in the clinical workflow every day~\cite{johnson2019mimic, demner2016preparing}. Jointly learning from images and raw text can support a leap in the quality of medical vision models by taking advantage of existing expert descriptions of the images.

Learning to extract \textit{useful} feature representations from training data is an essential objective of a deep learning model. The definition of \textit{usefulness} is task-specific~\cite{chen2018isolating, rifai2012disentangling, bojanowski2017unsupervised}. In this work, we aim to learn image representations that improve classification tasks, such as pathology detection, by making use of the rich information contained in the raw text that describe the findings in the image.

We exploit mutual information (MI) to learn useful image representations jointly with text. MI quantifies statistical dependencies between two random variables. Prior work has estimated and optimized MI across images for image registration~\cite{wells1996multi, maes1997multimodality}, and MI between images and image features for unsupervised learning~\cite{chen2016infogan, oord2018representation, hjelm2018learning}. Since the text usually describes image findings that are relevant for downstream image classification tasks, it is sensible to encourage the image and text representations to exhibit high MI.

\begin{figure}[!t]
\label{fig:ima_txt_eg}
\begin{minipage}{1.4in} 
  \caption{An example image-text pair (a chest radiograph and its associated radiology report). Each sentence describes the image findings in a particular region of the image. This figure is best viewed in color.}
\end{minipage} 
\hskip0.4in
\begin{minipage}{2.8in} 
\centerline{
\hfill
\includegraphics[width=3.3in]{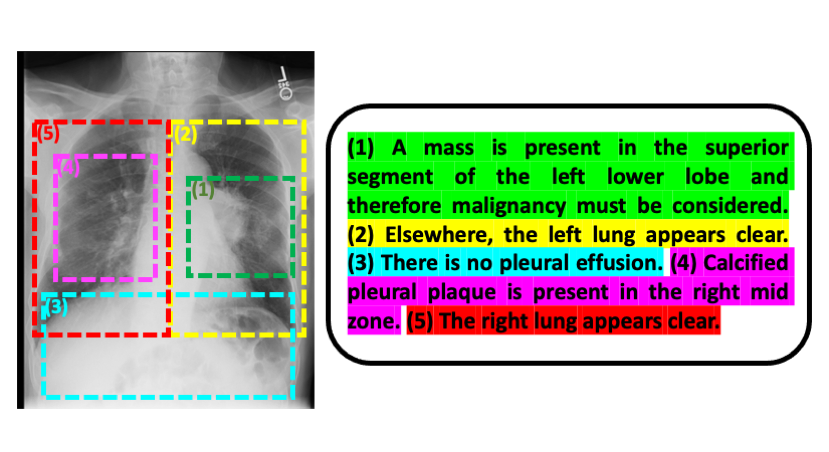}
\hfill
}
\end{minipage} 
\end{figure}

We propose to learn an image encoder and a text encoder by maximizing the MI of their resulting image and text representations. Moreover, we estimate and optimize the MI between local image features and sentence-level text representations. Fig.~\ref{fig:ima_txt_eg} shows an example image-text pair, where the image is a chest radiograph and the document is the associated radiology report~\cite{johnson2019mimic}. Each sentence in the report describes a local region in the image. A sentence is usually a minimal and complete semantic unit~\cite{zhang2020contrastive, reimers2019sentence}. The findings described in that semantic unit are usually captured in a local region of the image~\cite{harwath2018jointly}.

Prior work in image-text joint learning has leveraged image-based text generation as an auxiliary task during the image model training~\cite{wang2018tienet, xue2019improved, moradi2016cross}, or has blended image and text features for downstream inference tasks~\cite{moradi2018bimodal}. Other work has leveraged contrastive learning, an approach to maximize a lower bound on MI to learn image and text representations jointly~\cite{zhang2020contrastive, chauhan2020joint}. To the best of our knowledge, this work represents the first attempt to exploit the image spatial structure and sentence-level text features with MI maximization to learn image and text representations that are \textit{useful} for subsequent analysis of images. In our experimental results, we demonstrate that the maximization of local MI yields the greatest improvement in the downstream image classification tasks.

This paper is organized as follows. In Section~\ref{sec:methods}, we derive our approach for image-text representation learning by maximizing local MI. Section~\ref{sec:local_vs_global} discusses the theoretical motivation behind local mutual information. This is followed by empirical evaluation in Section~\ref{sec:experiments}, where we describe the implementation details of our algorithms in application to chest radiographs and radiology reports.

\section{Methods}
\label{sec:methods}
Let $x^{\text{I}}$ be an image and $x^{\text{R}}$ be the associated free text such as a radiology report or a pathology report that describes findings in the image. The objective is to learn useful  latent image representations~$z^{\text{I}}(x^{\text{I}})$ and text representations~$z^{\text{R}}(x^{\text{R}})$ from image-text data $\mathcal{X}=\{\text{x}_j\}_{j=1}^{N}$, where $\text{x}_j=(\text{x}^{\text{I}}_j, \text{x}^{\text{R}}_j)$. We construct an image encoder and a text encoder parameterized by~$\theta_\text{E}^\text{I}$ and~$\theta_\text{E}^\text{R}$, respectively, to generate the representations $z^{\text{I}}(x^{\text{I}}; \theta_\text{E}^\text{I})$ and $z^{\text{R}}(x^{\text{R}}; \theta_\text{E}^\text{R})$.

\paragraph{\textbf{Mutual Information Maximization.}} We seek such image and text encoders and learn their representations by maximizing MI between the image representation and the text representation:
\begin{align}
I(z^{\text{I}},z^{\text{R}}) \stackrel{\Delta}{=} \mathbb{E}_{p(z^{\text{I}},z^{\text{R}})} \left[\log \frac{p(z^{\text{I}},z^{\text{R}})}{p(z^{\text{I}})p(z^{\text{R}})}\right].
\label{eq:global_mi_def}
\end{align}
We employ MI as a statistical measure that captures dependency between images and text in the joint representation space. Maximizing MI between image and text representations is equivalent to maximizing the difference of the entropy and the conditional entropy of image representation given text: $I(z^{\text{I}},z^{\text{R}})=H(z^{\text{I}})-H(z^{\text{I}}|z^{\text{R}})$. This criterion encourages the model to learn feature representations where the information from one modality reduces the entropy of the other data modality, which is a better choice than solely minimizing the conditional entropy, where the image encoder could generate identical features for all data to achieve the conditional entropy minimum. 

\paragraph{\textbf{Stochastic Optimization of MI.}} Estimating mutual information between high-dimensional continuous variables from finite data samples is challenging. We leverage the recent advances that employ neural network discriminators for MI estimation and maximization~\cite{belghazi2018mine, oord2018representation, liao2020demi, song2019understanding}. The key idea is to construct a discriminator~$f(\text{z}_i^{\text{I}}, \text{z}_j^{\text{R}}; \theta_{\text{D}})$, parameterized by $\theta_{\text{D}}$, that estimates the likelihood (or the likelihood ratio) of whether a sample pair~$(\text{z}_i^{\text{I}}, \text{z}_j^{\text{R}})$ is sampled from the joint distribution~$p(z^{\text{I}},z^{\text{R}})$ or from the product of marginals~$p(z^{\text{I}})p(z^{\text{R}})$. The discriminator is commonly found by maximizing the lower bound of the MI approximated by the likelihood ratio in Eq.~(\ref{eq:global_mi_def})~\cite{belghazi2018mine, oord2018representation}.

We train the discriminator $f(\text{z}_i^{\text{I}}, \text{z}_j^{\text{R}}; \theta_{\text{D}})$ jointly with image and text encoders $z^{\text{I}}(x^{\text{I}}; \theta_\text{E}^\text{I})$ and $z^{\text{R}}(x^{\text{R}}; \theta_\text{E}^\text{R})$ via MI maximization:
\begin{align}
\hat{\theta}_\text{E}^\text{I}, \hat{\theta}_\text{E}^\text{R}, \hat{\theta}_\text{D} = \arg \max_{\theta_\text{E}^\text{I}, \theta_\text{E}^\text{R}, \theta_\text{D}} \hat{I}(z^{\text{I}}(x^{\text{I}}; \theta_\text{E}^\text{I}),z^{\text{R}}(x^{\text{R}}; \theta_\text{E}^\text{R}); \theta_{\text{D}}),
\label{eq:global_mi_approx}
\end{align}
where $\hat{I}(z^{\text{I}},z^{\text{R}}; \theta_{\text{D}})$ is a lower bound on $I(z^{\text{I}},z^{\text{R}})$. We consider two MI lower bounds: Mutual Information Neural Estimation (MINE)~\cite{belghazi2018mine} and Contrastive Predictive Coding (CPC)~\cite{oord2018representation}. In our experiments, we empirically show that our method is not sensitive to the choice of the lower bound. MINE estimates the MI lower bound by approximating the log likelihood ratio in Eq.~(\ref{eq:global_mi_def}), using the Donsker-Varadhan (DV) variational formula of the KL divergence between the joint distribution and the product of the marginals, which yields the lower bound
\begin{align}
 \hat{I}_{\theta_\text{E}^\text{I}, \theta_\text{E}^\text{R}, \theta_\text{D}}^{(\text{MINE})}(z^{\text{I}}, z^{\text{R}}) = \mathbb{E}_{p(z^{\text{I}},z^{\text{R}})} \left[f(z^{\text{I}},z^{\text{R}}; \theta_\text{D})\right] - \log \mathbb{E}_{p(z^{\text{I}})p(z^{\text{R}})} \left[e^{f(z^{\text{I}},z^{\text{R}}; \theta_\text{D})}\right].
\label{eq:global_mi_mine}
\end{align}
CPC computes the MI lower bound by approximating the likelihood of an image-text feature pair being sampled from the joint distribution over the product of marginals, which leads to the objective function
\begin{align}
 \hspace{-0.04in} \hat{I}_{\theta_\text{E}^\text{I}, \theta_\text{E}^\text{R}, \theta_\text{D}}^{(\text{CPC})}(z^{\text{I}}, z^{\text{R}}) = \mathbb{E}_{p(z^{\text{I}},z^{\text{R}})} \left[f(z^{\text{I}},z^{\text{R}}; \theta_\text{D})\right] - \mathbb{E}_{p(z^{\text{I}})p(z^{\text{R}})} \hspace{-0.05in} \left[\log \hspace{-0.08in} \sum_{\hat{\text{z}}_j^{\text{R}}\in z^{\text{R}}} \hspace{-0.08in} e^{f(z^{\text{I}}, \hat{\text{z}}_j^{\text{R}}; \theta_\text{D})}\right]\hspace{-0.04in}.
\label{eq:global_mi_cpc}
\end{align}

Both methods sample from the matched image-text pairs and from shuffled pairs (to approximate the product of marginals), and train the discriminator to differentiate between these two types of sample pairs.

\begin{figure*}[!b]
	\centering
	\includegraphics[width=1\linewidth]{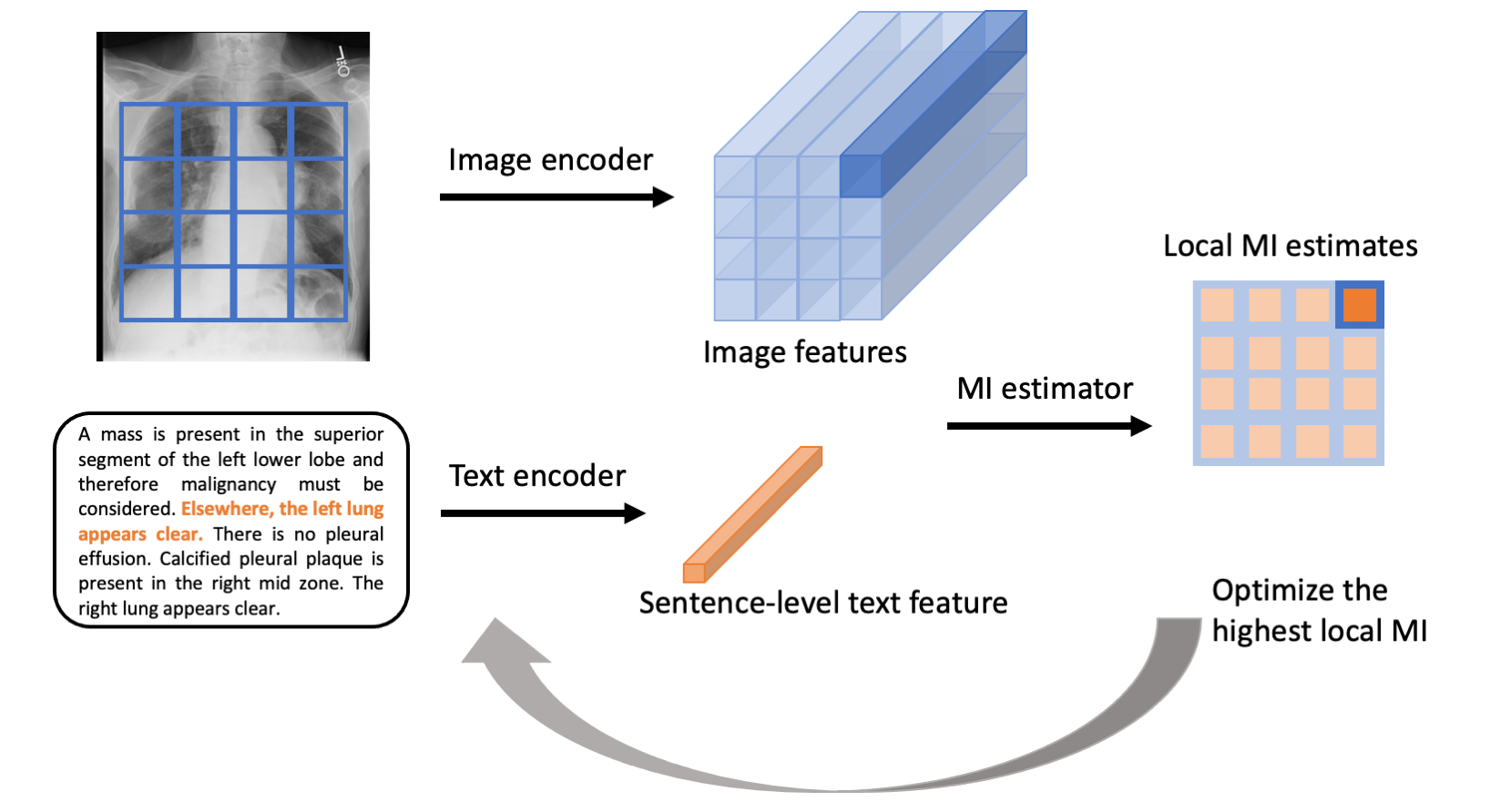}
	\caption{Local MI Maximization. First, we randomly select a sentence in the text and encode the sentence into a sentence-level feature. The corresponding image is encoded into a M$\times$M$\times$D feature block. We estimate the MI values between all local image features and the sentence feature. Note that the MI estimation needs shuffled image-text data, which is not illustrated in this diagram. We select the local image feature with the highest MI and update the image encoder, text encoder, and the MI discriminator such that the local MI between that image feature and the sentence feature is maximized.}
	\label{fig:algo_paradigm}
\end{figure*}

\paragraph{\textbf{Local MI Maximization.}} We propose to maximize MI between local features of images and sentence-level features from text. Given a sentence-level feature in the text, we estimate the MI values between all local image features and this sentence, select the image feature with the highest MI, and maximize the MI between that image feature and the sentence feature (Fig.~\ref{fig:algo_paradigm}). We train the image and text encoders, as well as the MI discriminator based on all the image-text data:
\begin{align}
\hat{\theta}_\text{E}^\text{I}, \hat{\theta}_\text{E}^\text{R}, \hat{\theta}_\text{D} = \arg \max_{\theta_\text{E}^\text{I}, \theta_\text{E}^\text{R}, \theta_\text{D}} \sum_{j} \sum_{m} \max_{n} \hat{I}(\text{z}_{j, (n)}^\text{I}, \text{z}_{j, (m)}^\text{R}),
\label{eq:local_mi_loss}
\end{align}
where $\text{z}_{j, (n)}^\text{I}$ is the $n$-th local feature in image $\text{x}_j^\text{I}$, and $\text{z}_{j, (m)}^\text{R}$ is the $m$-th sentence feature in text $\text{x}_j^\text{R}$. We use this \textit{one-way} maximum, because in image captioning, every sentence was written to describe some finding in the corresponding image. In contrast, not every region in the image has a related sentence in the text that describes it.

\section{Generative Model and Motivation}
\label{sec:local_vs_global}
\begin{figure}[!b]
	\centering
	\includegraphics[width=0.49\linewidth]{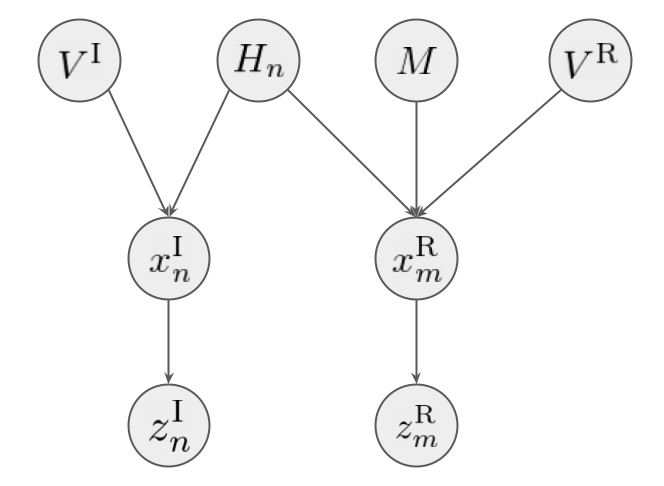}
	\caption{A conjectured generative model that describes how paired chest radiograph and radiology report are constructed and the underlying structural assumptions.}
	\label{fig:local_mi_graphical_model}
\end{figure}

To provide further insight into the theoretical motivation behind local mutual information, we describe a conjectured generative model for how paired chest radiograph and radiology report are constructed. As shown in Fig~\ref{fig:local_mi_graphical_model}, each local image region $x_n^\text{I}$ has a hidden variable $H_n$ that specifies the physiological processes and disease status in that region. This image region $x_n^\text{I}$ is generated by the hidden variable $H_n$ and another random variable $V^\text{I}$ that is independent of $H_n$ (e.g., the image acquisition protocol). The \textit{corresponding} sentence in the radiology report is generated by first choosing the sentence index $m$ (mapping from the image region index $n$ via $M$, i.e., $m=f(n;M)$) and then generated as a function of $H_n$ and another random variable $V^\text{R}$ that is independent of $H$ (e.g., the radiologist's training background).

The task we are interested in is to predict the hidden disease statuses~$\{H_n\}$ given an image~$x^I$. Therefore, it is sensible to learn an image feature representation $z^\text{I}$ that has high mutual information with $\{H_n\}$, i.e., $\sum_{n} I(z^\text{I}, H_n)$. $z^\text{I}$ is the concatenation of $z_n^\text{I}$ and $\bar{z_n^\text{I}}$, where the $z_n^\text{I}$ is the feature of the local image region generated from $H_n$ and $\bar{z_n^\text{I}}$ is the rest of the image features. Applying the chain rule of mutual information, we have:
\begin{align}
I(z^\text{I}, H_n) &= I(z_n^\text{I}, H_n)+I(\bar{z_n^\text{I}}, H_n|z^\text{I}) \\
&\geq I(z_n^\text{I}, H_n).
\label{eg:local_img_feature}
\end{align}
Since $I(z_n^\text{I}, H_n)$ is a lower bound to $I(z^\text{I}, H_n)$, we maximize $I(z_n^\text{I}, H_n)$. The challenge of learning such image feature representations is that we have limited labels for disease status. However, both the local image region and the \textit{corresponding} sentence in the report are generated by the same hidden disease status. Assuming $V^\text{I}$ and $V^\text{R}$ are independent, maximizing $I(z_n^\text{I}, z_m^\text{R})$ will likely lead to high $I(z_n^\text{I}, H_n)$, because $H_n$ is the only source of information shared by $z_n^\text{I}$ and $z_m^\text{R}$. Here we do the index mapping by selecting the sentence in the report that has the highest mutual information with $z_n^\text{I}$.

Therefore, conjecturing this generative model by making structural (conditional independence) assumptions of the image and report data results in our proposed local mutual information maximization approach. The local MI optimization is usually an easier task given its lower dimension and more training samples to discover useful representations. The utility of our strategy is supported by our experimental results.

\section{Experiments}
\label{sec:experiments}

\paragraph{\textbf{Data and Model Evaluation.}} We demonstrate our approach on the MIMIC-CXR dataset v2.0~\cite{johnson2019mimic} that includes around 250K frontal-view chest radiographs with their associated radiology reports. We evaluate our representation learning methods on two downstream classification tasks: 
\begin{itemize}
    \item \textbf{Pathology9}. Detecting 9 pathologies from the chest radiographs against the labels that were extracted from the corresponding radiology reports using a radiology report labeler CheXpert~\cite{irvin2019chexpert, johnson2019mimicjpg, johnson2019mimicjpgphysio}. Note that there are 14 findings available in the repository~\cite{johnson2019mimicjpgphysio}. We only train and evaluate 9 for which there are more than around 100 images available in the test set.
    \item \textbf{EdemaSeverity}. Assessing pulmonary edema severity from chest radiographs against the labels that were annotated by radiologists on the images~\cite{horng2021deep, liao2021edemalabels, liao2019semi, wang2019pulmonary}. The severity level ranges from 0 to 3 with a higher score indicating higher risk. 
\end{itemize}

The test sets provided in MIMIC-CXR with CheXpert labels~\cite{johnson2019mimicjpgphysio} and with edema severity labels~\cite{liao2021edemalabels} are used to evaluate our methods. The patients that are in either of the two test sets are excluded from the model training. Table~\ref{table: data_size} summarizes the size of the (labeled) training data and test data.

\begin{table}[!b]
\centering

\begin{tabular}{|p{1.3cm}|p{2.4cm}|p{2cm}|p{2.1cm}|p{1.6cm}|p{2.3cm}|}
\hline
-- & Support Devices & Cardiomegaly & Consolidation & Edema & Lung Opacity \\ \hline
training & 76,492 & 65,129 & 20,074 & 56,203 & 58,105  \\\hline
test & 286 & 404 & 95 & 373 & 318  \\\hline
\end{tabular}

\begin{tabular}{|p{1.3cm}|p{2.4cm}|p{2cm}|p{2.1cm}|p{1.6cm}|p{2.3cm}|}
\hline
-- & Pleural Effusion & Pneumonia & Pneumothorax & Atelectasis & Edema Severity \\ \hline
training & 86,871 & 43,951 & 56,472 & 50,416 & 7,066  \\\hline
test & 451 & 195 & 191 & 262 & 141  \\\hline
\end{tabular}

\vspace{0.1in}
\caption{The number of images in the (labeled) training sets and the test sets.}
\label{table: data_size}
  
\end{table}

\paragraph{\textbf{Experimental Design.}}
Our goal is to learn representations that are useful for downstream classification tasks. Therefore, we use a fully supervised image model trained on the chest radiographs with available training labels as our benchmark. We compare two ways to use our image representations when \textit{re-training} the image classifier: 1) freezing the image encoder; 2) fine-tuning the image encoder. In either case, the image encoder followed by a classifier is trained on the same training set that is used to train the fully supervised image model. 

We compare our MI maximization approach on local features with the global MI maximization. We test both MINE~\cite{belghazi2018mine} and CPC~\cite{oord2018representation} as MI estimators. To summarize, we evaluate the variants of our model and training regimes as follows:
\begin{itemize}
    \item \textbf{image-only-supervised}: An image-only model trained on the training data provided in~\cite{johnson2019mimicjpgphysio, liao2021edemalabels}.
    \item \textbf{global-mi-mine}, \textbf{global-mi-cpc}: Representation learning on the chest radiographs and the radiology reports using global MI maximization.
    \begin{itemize}
        \item \textbf{encoder-frozen}, \textbf{encoder-tuned}: Once representation learning is completed, the image encoder followed by a classifier is \textit{re-trained} on the labeled training image data, with the encoder frozen or fine-tuned.  
    \end{itemize}
    \item \textbf{local-mi-mine}, \textbf{local-mi-cpc}: Representation learning using local MI maximization in Eq.~(\ref{eq:local_mi_loss}).
    \begin{itemize}
        \item \textbf{encoder-frozen}, \textbf{encoder-tuned}: The resulting image encoder followed by a classifier is \textit{re-trained}, with the encoder frozen or fine-tuned.
    \end{itemize}
\end{itemize}

At the image model training or \textit{re-training} time, all variants are trained on the same training sets. Note that the \textbf{local-mi} approach makes use of lower level image features. To make the \textbf{encoder-frozen} experiments comparable between \textbf{local-mi} and \textbf{global-mi}, we only freeze the same lower level feature extractor in both encoders. 

\paragraph{\textbf{Implementation Details.}}
Chest radiographs are downsampled to 256$\times$256. We use a 5-block resnet~\cite{he2016deep} as the image encoder in the local MI approach and the image feature representation~$z^{\text{I}}$ is 16$\times$512 (4$\times$4$\times$512) feature vectors. We use a 6-block resnet as the image encoder for the global MI maximization, where the image representation~$z^{\text{I}}$ from this encoder is a 768-dimensional feature vector. We use the clinical BERT model \cite{alsentzer2019publicly} as the text encoder for both report-level and sentence-level feature extraction. The \texttt{[CLS]} token is used as the text feature~$z^\text{R}$, which is a 768-dimensional vector. The MI discriminator for both MINE and CPC is a 1280$\rightarrow$1024$\rightarrow$512$\rightarrow$1 multilayer perceptron to estimate local MI and a 1536$\rightarrow$1024$\rightarrow$512$\rightarrow$1 multilayer perceptron to estimate global MI. The image feature and the text feature are concatenated to construct the input for the discriminator for MI estimation. The image models in all training variants at the image training or \textit{re-training} time have the same architecture (6-block resnet followed by a fully connected layer). 

The AdamW~\cite{wolf2019huggingface} optimizer is employed for the BERT encoder and the Adam~\cite{kingma2014adam} optimizer is used for the other parts of the model. The initial learning rate is $5{\cdot}10^{-4}$. The representation learning phase is trained for 5 epochs and the image model \textit{re-training} phase is trained for 50 epochs. The fully supervised image model is trained for 100 epochs. Data augmentation including random rotation, translation, and cropping is performed on the images during training.

\begin{table}[!h]
\centering

\begin{tabular}{|P{1.9cm}|c || p{1cm}|p{1cm} || p{1cm} | p{1cm} || p{1cm} | p{1cm} ||}
\hline
Method & \textit{Re-train} Encoder? & \multicolumn{2}{ c ||}{Level 0 vs 1,2,3} & \multicolumn{2}{c||}{Level 0,1 vs 2,3} & \multicolumn{2}{ c ||}{Level 0,1,2 vs 3}\\ \hline
-- & -- & CPC & MINE & CPC & MINE & CPC & MINE\\ \hline
\textbf{image-only} & N/A & \multicolumn{2}{c ||}{0.80} & \multicolumn{2}{c ||}{0.71} & \multicolumn{2}{c ||}{0.90}  \\\hline
\textbf{global-mi} & \textbf{frozen} & 0.81 & 0.83 & 0.77 & 0.78 & 0.93 & 0.89\\\hline
\textbf{global-mi} & \textbf{tuned} & 0.81 & 0.82 & 0.79 & 0.81 & 0.93 & 0.93 \\\hline
\textbf{local-mi}  & \textbf{frozen} & 0.77 & 0.76 & 0.72 & 0.76 & 0.75 & 0.86 \\\hline
\textbf{local-mi}  & \textbf{tuned} & \textbf{0.87} & 0.83 & 0.83 & \textbf{0.85} & \textbf{0.97} & 0.93 \\\hline
\end{tabular}
\vspace{0.1in}
\caption{The AUCs on the \textbf{EdemaSeverity} ordinal regression task. The average AUC score of \textbf{tuned} \textbf{local-mi} is 0.88 ($\pm$0.05); The average AUC score of \textbf{tuned} \textbf{global-mi} is 0.85 ($\pm$0.06).}
\label{table: auc_edema_severity}

\end{table}

\vspace{-0.5in}

\begin{table}[!h]
\centering

\begin{tabular}{|P{1.9cm}|c || p{1cm}|p{1cm} || p{1cm} | p{1cm} || p{1.2cm} | p{1.2cm} ||}
\hline
Method & \textit{Re-train} Encoder? & \multicolumn{2}{ c ||}{Atelectasis} & \multicolumn{2}{c||}{Cardiomegaly} & \multicolumn{2}{ c ||}{Consolidation} \\ \hline
-- & -- & CPC & MINE & CPC & MINE & CPC & MINE \\ \hline
\textbf{image-only} & N/A & \multicolumn{2}{c ||}{0.76} & \multicolumn{2}{c ||}{0.71} & \multicolumn{2}{c ||}{0.78}  \\\hline
\textbf{global-mi} & \textbf{frozen} & 0.65 & 0.63 & 0.79 & 0.79 & 0.67 & 0.65\\\hline
\textbf{global-mi} & \textbf{tuned} & 0.74 & 0.77 & 0.81 & 0.81 & 0.81 & 0.82 \\\hline
\textbf{local-mi}  & \textbf{frozen} & 0.74 & 0.61 & 0.73 & 0.77 & 0.65 & 0.65 \\\hline
\textbf{local-mi}  & \textbf{tuned} & 0.73 & \textbf{0.86} & 0.82  & \textbf{0.84} & \textbf{0.83} & \textbf{0.83}  \\\hline
\end{tabular}

\begin{tabular}{|P{1.9cm}|P{2.6cm}|| p{1cm}|p{1cm} || p{1cm} | p{1cm} || p{1.2cm} | p{1.2cm} ||}
\hline 
-- & -- & \multicolumn{2}{ c ||}{Edema} & \multicolumn{2}{c||}{Lung Opacity} & \multicolumn{2}{ c ||}{Pleural Effusion} \\ \hline
-- & -- & CPC & MINE & CPC & MINE & CPC & MINE \\ \hline
\textbf{image-only} & N/A & \multicolumn{2}{c ||}{\textbf{0.89}} & \multicolumn{2}{c ||}{0.86} & \multicolumn{2}{c ||}{0.69}  \\\hline
\textbf{global-mi} & \textbf{frozen} & 0.81 & 0.81 & 0.69 & 0.68 & 0.74 & 0.74\\\hline
\textbf{global-mi} & \textbf{tuned} & 0.87 & 0.88 & 0.83 & 0.84 & 0.90 & 0.90 \\\hline
\textbf{local-mi}  & \textbf{frozen} & 0.78 & 0.80 & 0.66 & 0.69 & 0.69 & 0.72\\\hline
\textbf{local-mi}  & \textbf{tuned} & \textbf{0.89} & \textbf{0.89} & 0.82 & \textbf{0.88} & \textbf{0.92} & \textbf{0.92}\\\hline
\end{tabular}

\begin{tabular}{|P{1.9cm}|P{2.6cm}|| p{1cm}|p{1cm} || p{1cm} | p{1cm} || p{1.2cm} | p{1.2cm} ||}
\hline 
-- & -- & \multicolumn{2}{ c ||}{Pneumonia} & \multicolumn{2}{c||}{Pneumothorax} & \multicolumn{2}{ c ||}{Support Devices} \\ \hline
-- & -- & CPC & MINE & CPC & MINE & CPC & MINE \\ \hline
\textbf{image-only} & N/A & \multicolumn{2}{c ||}{0.75} & \multicolumn{2}{c ||}{0.65} & \multicolumn{2}{c ||}{0.72}  \\\hline
\textbf{global-mi} & \textbf{frozen} & 0.71 & 0.70 & 0.65 & 0.66 & 0.70 & 0.68\\\hline
\textbf{global-mi} & \textbf{tuned} & 0.75 & 0.76 & 0.75 & 0.77 & 0.77 & 0.79 \\\hline
\textbf{local-mi}  & \textbf{frozen} & 0.61 & 0.66 & 0.70 & 0.67 & 0.72 & 0.74\\\hline
\textbf{local-mi}  & \textbf{tuned} & 0.78 & \textbf{0.79} & \textbf{0.79} & 0.76 & \textbf{0.87} & 0.81 \\\hline
\end{tabular}

\vspace{0.1in}
\caption{The AUCs on the \textbf{Pathology9} binary classification tasks. The average AUC score of \textbf{tuned} \textbf{local-mi} is 0.84 ($\pm$0.05); The average AUC score of \textbf{tuned} \textbf{global-mi} is 0.81 ($\pm$0.05).}
\label{table: auc_pathology9}
  
\end{table}

\vspace{-0.4in}

\paragraph{\textbf{Results.}} In Table~\ref{table: auc_edema_severity} and Table~\ref{table: auc_pathology9}, we present the area under the receiver operating characteristic curve (AUC) statistics for the variants of our algorithms on the \textbf{EdemaSeverity} classification task and the \textbf{Pathology9} binary classification tasks. For most classification tasks, the local MI approach with encoder tuning performs the best and significantly improves the performance over solely supervised learning on labeled images. The local MI approach brings in noteworthy improvement compared to global MI. Both CPC and MINE perform similar in most tasks. Remarkably, the classification results from the frozen encoders approach the fully supervised learning results in many tasks, suggesting that the unsupervised learning captures useful features for image classification tasks even before supervision is provided.

The local MI offers substantial improvement in performance when the features are fine-tuned with the downstream model, while its performance is comparable with global MI if the features are frozen for the subsequent classification. In our experiments, training jointly with the downstream classifier (fine-tuning) typically improves performance of all tasks, with greater benefits for local MI. This suggests that local MI yields more flexible representations that adjust better for the downstream task. Our results are also supported by the analysis in Section~\ref{sec:local_vs_global} that shows certain structural assumptions lead to the local MI approach, which is easier to discover useful representations due to its lower dimension and more training samples.

\section{Conclusion}
\label{sec:conclusion}
In this paper, we proposed a multimodal representation learning framework for images and text by maximizing the mutual information between their local features. The advantages of the local MI approach are tri-fold: 1) better fit to image-text structure: each sentence is typically a minimal and complete semantic unit that describes a local image region (Fig.~\ref{fig:ima_txt_eg}) and therefore learning at the level of sentences and local image regions is more efficient than learning global descriptors; 2) better optimization landscape: the dimensionality of the representation is lower and every training image-report pair provides more samples of image-text descriptor pairs; 3) better representation fit to downstream tasks: as demonstrated in prior work, image classification usually relies on local features (e.g., pleural effusion detection based on the appearance of the region below the lungs)~\cite{hjelm2018learning} and thus by learning local representations local MI improves classification performance.

By encouraging sentence-level features in the text to exhibit high MI with local image features, the image encoder learns to extract \textit{useful} feature representations for subsequent image analysis. We provided further insight into local MI by showing that, under a Markov condition, maximizing local MI is equivalent to maximizing global MI. Our experimental results demonstrate that the local MI approach offers the greatest improvement for the downstream image classification tasks, and is not sensitive to the choice of the MI estimator. 

\subsubsection*{Acknowledgments.} This work was supported in part by NIH NIBIB NAC P41EB015902, Wistron, IBM Watson, MIT Deshpande Center, MIT J-Clinic, MIT Lincoln Lab, and US Air Force. 

%\small
\bibliographystyle{splncs04}
\bibliography{paper2195}





% LNCS Homepage, \url{http://www.springer.com/lncs}. Last accessed 4
% \end{thebibliography}
\end{document}